\documentclass[%
 aip,
 amsmath,amssymb,
 reprint,%
]{revtex4-1M}

\usepackage{graphicx}
\usepackage{dcolumn}
\usepackage{bm}
\usepackage[utf8]{inputenc}
\usepackage[T1]{fontenc}
\usepackage{mathptmx}
\usepackage{verbatim}
\usepackage{xcolor}
\usepackage{epstopdf}
\AppendGraphicsExtensions{.tif}

\begin{document}

\preprint{APL-Mesoscopic Magnetic Systems}

\title[Skyrmion Logic Clocked via Voltage Controlled Magnetic Anisotropy]{Skyrmion Logic Clocked via Voltage Controlled Magnetic Anisotropy}

\author{Benjamin W. Walker}
\affiliation{Department of Electrical and Computer Engineering, The University of Texas at Dallas, Richardson, TX 75080, USA.}
\author{Can Cui}
\affiliation{Department of Electrical and Computer Engineering, The University of Texas at Austin, Austin, TX 78712, USA.}
\author{Felipe Garcia-Sanchez}
\affiliation{Departamento de Física Aplicada, Universidad de Salamanca, 37008 Salamanca, Spain.}
\author{Jean Anne C. Incorvia}
\affiliation{Department of Electrical and Computer Engineering, The University of Texas at Austin, Austin, TX 78712, USA.}
\author{Xuan Hu}
\affiliation{Department of Electrical and Computer Engineering, The University of Texas at Dallas, Richardson, TX 75080, USA.}
\author{Joseph S. Friedman}
\affiliation{Department of Electrical and Computer Engineering, The University of Texas at Dallas, Richardson, TX 75080, USA.}

\date{\today}

\begin{abstract}
Magnetic skyrmions are exciting candidates for energy-efficient computing due to their non-volatility, detectability, and mobility. A recent proposal within the paradigm of reversible computing enables large-scale circuits composed of directly-cascaded skyrmion logic gates, but it is limited by the manufacturing difficulty and energy costs associated with the use of notches for skyrmion synchronization. To overcome these challenges, we therefore propose a skyrmion logic synchronized via modulation of voltage-controlled magnetic anisotropy (VCMA). In addition to demonstrating the principle of VCMA synchronization through micromagnetic simulations, we also quantify the impacts of current density, skyrmion velocity, and anisotropy barrier height on skyrmion motion. Further micromagnetic results demonstrate the feasibility of cascaded logic circuits in which VCMA synchronizers enable clocking and pipelining, illustrating a feasible pathway toward energy-efficient large-scale computing systems based on magnetic skyrmions.
\end{abstract}

\maketitle

\section{Introduction}

The non-volatility and rich physical interactions of magnetic skyrmions have inspired interest in their application to logical computing. Magnetic skyrmions are topologically-stable quasiparticles resulting from the Dzyaloshinskii-Moriya interaction (DMI),\cite{Everschor-Sitte2018} and can be propelled along a track via an applied current. \cite{Sampaio2013,Tomasello2014} Their non-volatilility, thermal stability, and energy-efficient motion have led to proposals for their use in racetrack memory systems. \cite{Tomasello2014, Kang2018, Behera2020}

Several approaches for skyrmion logic have also been proposed, though the early work in this field did not provide a scalable technique for cascading logic gates without requiring external control, modulation, or amplification circuitry. \cite{SZhang2015,XZhang2015,XZhang2015(2),xing2016,Luo2018,XZhang2015(3)} We therefore recently proposed a scalable paradigm for skyrmion logic based on reversible computing and conservative logic in which skyrmion logic gates can be directly cascaded without any external circuitry.\cite{Chauwin2019}

Skyrmion synchronization is critical to the functionality of scalable skyrmion logic systems, as the key skyrmion-skyrmion interactions require the simultaneous arrival of skyrmions from different parts of the circuit. Given the differing lengths of the skyrmion tracks and the inevitable thermal-induced variations of skyrmion velocity, clocked synchronization elements must be included to ensure proper logical functionality. The proposed use of notch synchronizers\cite{Chauwin2019} constrict skyrmion motion such that at normal operating current densities, skyrmions are too large to fit through the constriction. The skyrmion logic system is clocked with periodic increases to the current density, causing the skyrmion radii to temporarily decrease and allow skyrmions to pass through the notch constrictions. However, these notch synchronizers are difficult to precisely manufacture and the high current densities lead to skyrmion annihilation and high power dissipation.

To circumvent the challenges of notch synchronizers, this paper proposes the clocking of skyrmion logic systems via voltage-controlled magnetic anisotropy (VCMA). By modulating VCMA within the skyrmion track,\cite{XZhang2015(2)} skyrmions can be synchronized with less energy, higher reliability, and greater robustness to device imperfection. We illustrate the functionality of the proposed VCMA synchronizers through micromagnetic simulation, and quantify the anisotropy barrier needed to pin skyrmions as a function of current density and skyrmion velocity. Furthermore, we discuss fabrication approaches and demonstrate through micromagnetic simulation that these VCMA synchronizers can be integrated in large cascaded circuits with low power dissipation and high robustness.

\section{Skyrmion Logic and VCMA Fundamentals}

The skyrmion logic system of Ref. 12 leverages the rich physics of magnetic skyrmions to perform logical operations. As skyrmion synchronization is critical to the operation of these logic gates, notch synchronizers were suggested to ensure the correct functionality of cascaded circuits. As this notch-based synchronization is both inefficient and error-prone, VCMA is an intriguing approach to efficiently pin skyrmions.
\vspace{-1em}

\begin{figure}[!t]
\includegraphics[trim = 0 105 175 0, clip, width=1\columnwidth]{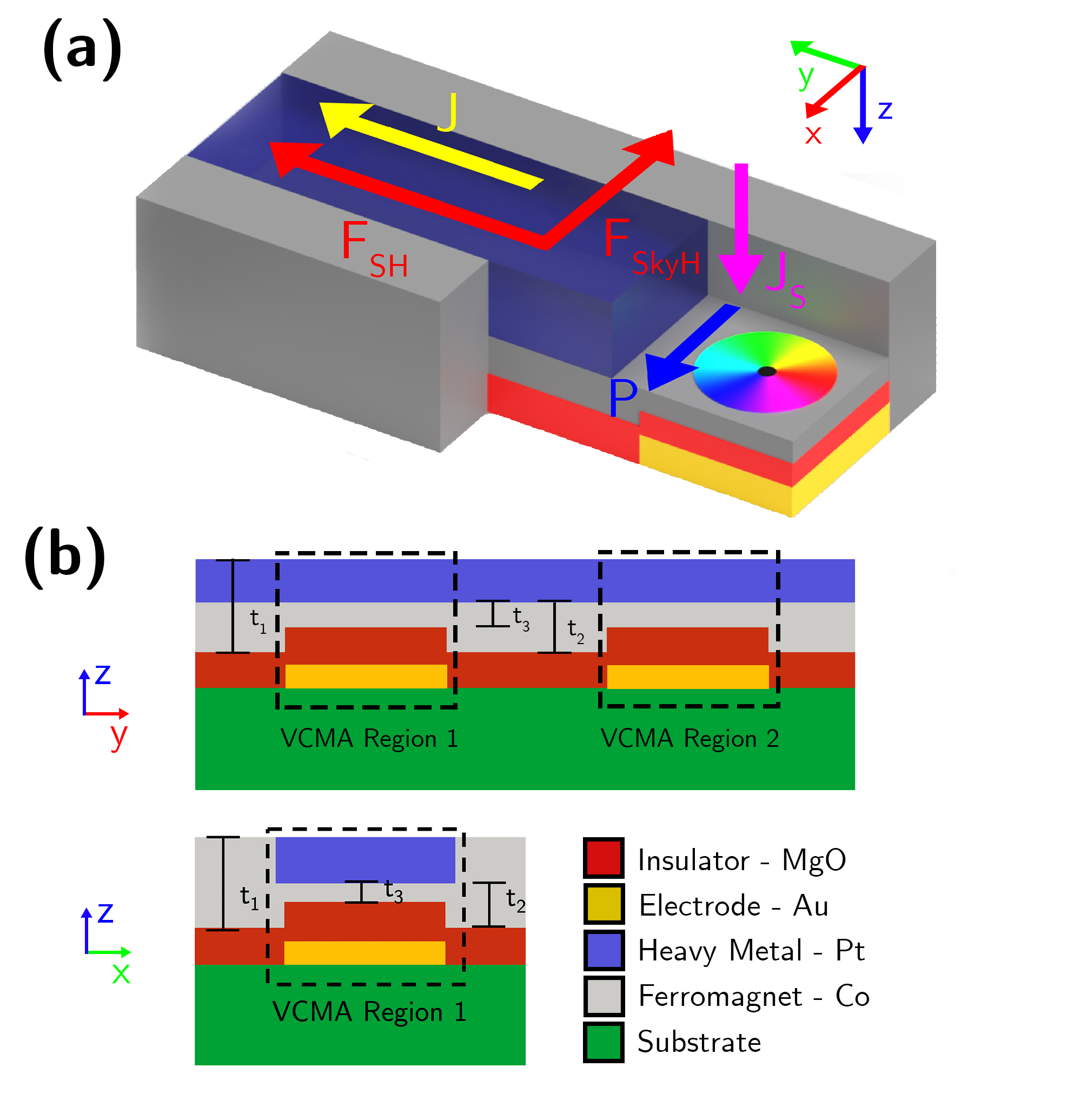}
\caption{VCMA synchronizer structure. (a) A Néel skyrmion (colored circle) exists at the interface between a heavy metal and a ferromagnet with polarization $P$. Injected electric current in the $+y$ direction ($J$) induces a spin current ($J_{S}$) in the $+z$ direction via the spin-Hall effect. This spin current produces a $+y$-directed spin-Hall force ($F_{SH}$), propelling the skyrmion in the $+y$ direction. Voltage applied to the electrode creates an electric field through the insulator that alters the magnetic anisotropy of the interface. (b) Synchronizer cross sections in the $yz$ and $xz$ planes. 
}
\label{fig:schematic}
\end{figure}

\subsection{Skyrmion Kinetics}

The skyrmion logic system of Ref. 12 utilizes four distinct skyrmion phenomena. As illustrated in Fig. \ref{fig:schematic}(a), current injected in the $+y$ direction through a ferromagnetic track creates a spin current in the $+z$ direction via the spin-Hall effect. This spin current causes a first force that pushes the skyrmion in the $+y$ direction. \cite{Sampaio2013, Tomasello2014} $y$-directed skyrmion motion results in a $-x$-directed deviation from linear $y$-directed motion via a second force due to the skyrmion-Hall effect (SkHE),\cite{Wang2018} and is countered by the third force, edge repulsion, to prevent skyrmion annihilation. The fourth phenomenon is the skyrmion-skyrmion repulsion between two skyrmions in close proximity.\cite{Wang2017}\vspace{-1em}

\subsection{Reversible Skyrmion Logic Gates}
The skyrmion logic system of Ref. 12 enables a range of skyrmion logic gates and encodes binary `1' and `0' values through the presence or absence of a skyrmion, respectively. Constant injected current drives skyrmions along tracks towards logic gate junctions, where interplay between the skyrmion-Hall effect and skyrmion-skyrmion repulsion causes the skyrmion trajectories to depend on the binary states of the inputs. As illustrated by the example AND/OR gate of Supplemental Material Fig. S1 and Video S1, the logical outputs are encoded in the output tracks to which the skyrmions are driven.\cite{Chauwin2019}

In these reversible skyrmion logic systems, all information is conserved, theoretically permitting energy dissipation below the Landauer limit. Relatedly, skyrmions are never destroyed and therefore need not be nucleated, as they can be indefinitely cascaded and reused.\cite{Chauwin2019} The final outputs of this skyrmion logic system can be read via magnetic tunnel junctions. \cite{hamamoto2018}

\subsection{Notch Synchronizer}

Skyrmion logic gates are cascaded by using the output skyrmions from one gate as the input skyrmions of other gates. As the gates require skyrmion synchronization to ensure proper skyrmion-skyrmion repulsion within the junctions, Ref. 12 proposed a notch synchronizer between cascaded logic gates (Supplementary Material Fig. S2 and Video S2). The notch structure pins skyrmions when normal low current density is applied, as the skyrmion radius is too large. By periodically providing a larger current pulse, the skyrmion radii shrink such that they depin and pass through the constriction.

By placing multiple synchronizers in a skyrmion logic circuit, a global clocking current pulse can force the skyrmions to interact at the track junctions, thereby ensuring correct logical operation despite thermal variations in velocity and differing trajectory lengths. However, this large current pulse is energetically expensive and can cause rapid skyrmion acceleration that can lead to skyrmion annihilation. It is also difficult to fabricate the notch structure with sufficient precision to ensure proper synchronization, and it is therefore worthwhile to consider alternative approaches.

\subsection{Voltage-Controlled Magnetic Anisotropy}

VCMA is a technique for varying the uniaxial anisotropy of a magnetic material via an applied electric field. A voltage applied across the ferromagnetic track changes its electron density of states which, in turn, changes the perpendicular magnetic anisotropy and magnetic coercivity.\cite{Wang2016} The relationship between applied voltage and change in anisotropy is predominantly linear, following the expression $K_{uv} = K_u + \vartheta V_b $. Where $K_{uv}$ is the resultant anisotropy, $K_u$ is the original anisotropy, $V_b$ is the bias voltage, and $\vartheta$ is a material-dependent coefficient ranging for most materials between $20 \frac{\mu J}{m^2}/\frac{V}{nm}$ and $100 \frac{\mu J}{m^2}/\frac{V}{nm}$ for Fe/Vacuum and Fe/MgO interfaces respectively.\cite{Wang2012}

When applied to the skyrmion track, VCMA can be used to pin a skyrmion via an anisotropy gradient.\cite{XZhang2015(2)} An applied positive voltage increases the anisotropy of the VCMA region which creates a positive energy barrier for a skyrmion entering the region and a negative energy barrier for a skyrmion leaving the region. For a negative voltage, the respective energy barriers are reversed.\cite{Wang2016} Skyrmions without sufficient driving force or velocity will be unable to pass, and will be pinned at the interface between the high and low anisotropy regions.

\section{VCMA Synchronizer}

We propose the synchronization of skyrmion logic systems by periodically switching skyrmion pinning via modulation of VCMA, enabling logic circuits that are more reliable and energy-efficient than can be achieved with notch synchronizers. These VCMA synchronizers can be readily fabricated with existing technology, and their functionality has been demonstrated via micromagnetic simulation.

\begin{figure}[t]

\includegraphics[width=0.95\columnwidth]{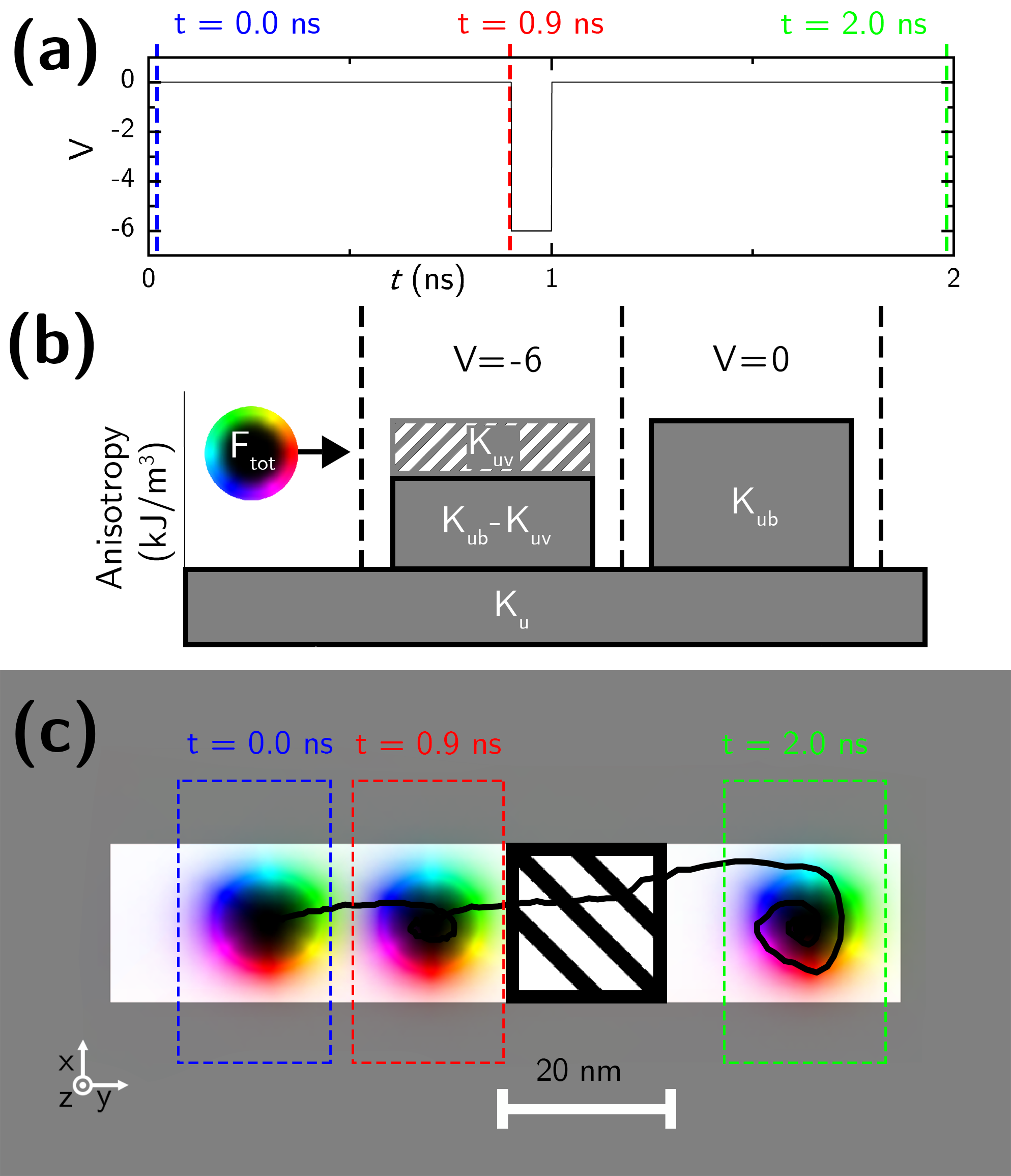}
\caption{VCMA synchronizer with a constant injected current of $5\times 10^{10} J/m^2$ in the $+y$ direction. (a) Voltage waveform applied to the VCMA region as a function of time. (b) Depiction of anisotropy values for various applied voltages. The skyrmion is able to pass when $F_{tot} > K_{ub}-K_{uv}$. (c) Micromagnetic simulation results showing skyrmion trajectory (black line) and locations over time. VCMA region is represented by the dashed square. The skyrmion is pinned at $t = 0.9$ ns by the positive anisotropy barrier of the VCMA region. At $t = 1.0$ ns, the clock voltage reduces the anisotropy barrier and the skyrmion depins. Simulations performed via mumax3 \cite{Mumax} with damping constant $\alpha = 0.1$, linear perpendicular anisotropy $K = 6\times 10^5 J/m^{2}$, interfacial DMI strength $D = 3\times 10^{-3} {J}/{m^2}$, exchange stiffness $A = 1.5 \times 10^{-11} J/m$, and saturation magnetization $Ms = 5.8 \times 10^5 A/m$ chosen to match the properties of Co/Pt films. Simulations were performed with a cell size of $1 \times 1 \times 0.4 \hspace{4px} nm^3$. The combined track thickness was 1.6 nm. Subsequent simulations all use identical parameters.\vspace{-1em}}
\label{fig:operation}
\end{figure}

\subsection{Structure}

Fig. \ref{fig:schematic}(a) and (b) shows the top and cross-sectional views of a VCMA synchronizer, respectively. The bottom electrode (Au) is coupled to the ferromagnetic racetrack (Co) via the insulator (MgO), thus allowing an electric field to be applied to the racetrack to modulate its anisotropy. $t_1$ and $t_2$ represent the track thickness and track wall, respectively. In order for skyrmions to be pinned at the VCMA region in the absence of an applied electric field, a preexisting anisotropy bias is introduced to the VCMA region by locally reducing the thickness of the track from $t_2$ to $t_3$.\cite{Carcia1988,Metaxas2007}

Experimentally, the proposed VCMA synchronizer can be fabricated using standard semiconductor processing techniques. Bottom electrodes and cobalt racetrack regions with different thicknesses ($t_1$, $t_2$, $t_3$) can be created with metal deposition, electron-beam lithography/photolithography and Ar ion milling. Chemical-mechanical polishing can be used to ensure the smoothness of the skyrmion racetrack.

\subsection{Operation}

The operation of the VCMA skyrmion synchronizer is demonstrated in the micromagnetic simulation results of Fig. \ref{fig:operation} and Video 1. A global electric current applied in the $+y$ direction accelerates the skyrmion in the $+y$ direction, though it is pinned by the VCMA region when a clock voltage of $0 V$ is applied. At $t = 0.9$ ns, a clock pulse of $-6 V$ temporarily lowers the anisotropy, depinning the skyrmion. By including numerous such VCMA regions controlled by the same global clock, skyrmions can be synchronized within large-scale cascaded skyrmion logic circuits to ensure proper skyrmion-skyrmion repulsion with the logic gates and correct logical functionality.

90\% of the clock cycle remains at zero volts, which reduces the amount of leaked charge, improving static power dissipation. Additionally, as it pins skyrmions without applied voltage, the system preserves memory without power. As the VCMA synchronizer avoids the high current densities of the notch-based design, it is more dynamically power efficient. Further, as high current densities often destroy skyrmions, the VCMA synchronizer is also more reliable.

\begin{figure}[t]
\includegraphics[trim = 5 0 10 0, clip, width=\columnwidth]{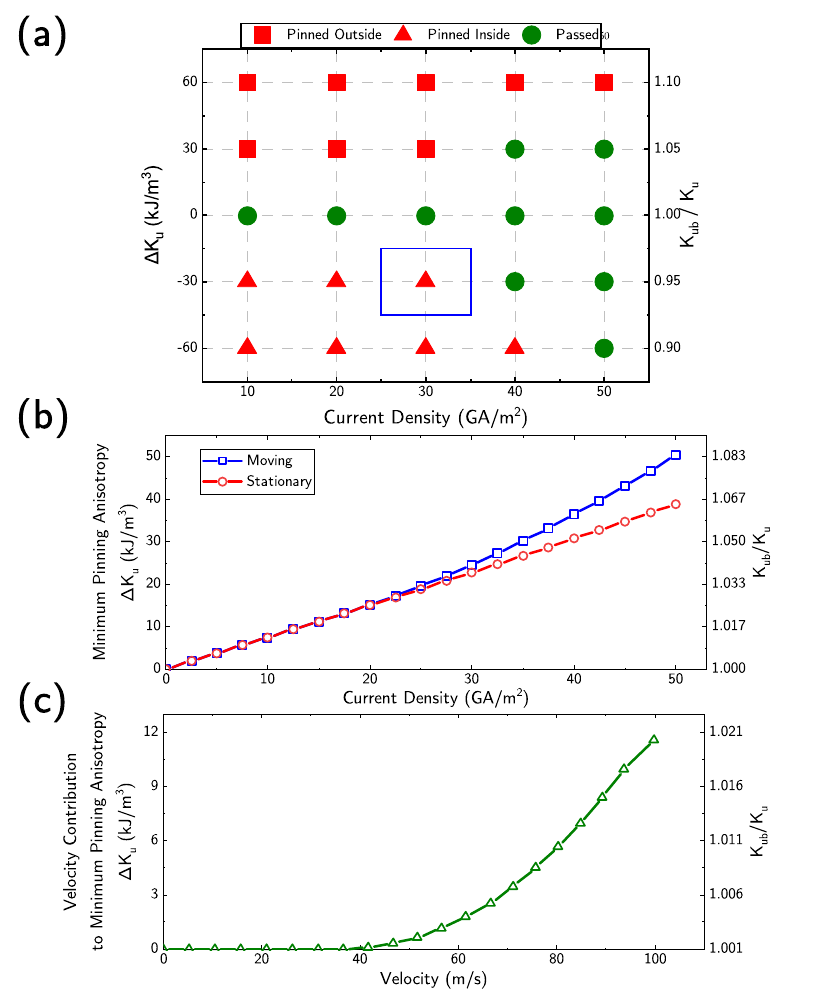}
\vspace{-2em}
\caption{
(a) Conditions under which a skyrmion in a region with anisotropy $K_u$ is able to traverse a region with anisotropy $K_{ub}$ are indicated with green circles, while skyrmions that get pinned outside the $K_{ub}$ region before entering and skyrmions that get pinned within the $K_{ub}$ region are indicated with red squares and triangles, respectively. Micromagnetic simulations were performed under a variety of current densities for skyrmions that started while both stationary and moving at terminal velocity. The blue square indicates the only condition under which the initial skyrmion velocity impacted the result; for this condition, the skyrmion that was initially stationary is pinned, while the skyrmion that is initially moving at its terminal velocity passes the anisotropy barrier. (b) The minimum anisotropy barrier required $(\Delta K_u)$ to pin a skyrmion at a VCMA region for skyrmions moving at terminal velocity (blue squares) and for initially-stationary skyrmions (red circles) as a function of current density. (c) The contribution of velocity on the minimum pinning anisotropy as a function of velocity, computed via the difference in the minimum pinning anisotropies of moving and stationary skyrmions.} 
\label{fig:anisotropy_analysis}
\end{figure}

\section{VCMA Synchronizer Analysis and Design}

By quantifying the interaction between a moving skyrmion and the VCMA region, we can optimize the efficiency, accuracy, and speed of VCMA-clocked skyrmion logic systems. This analysis reveals that both the injected current and the velocity of the skyrmion contribute to the total effective driving force. Governed by the Thiele equation, this force determines whether a skyrmion can overcome a specified anisotropy barrier, and therefore is crucial for selecting the PMA of the VCMA region.

\subsection{Anisotropy Barrier}

The ability of skyrmions to traverse a VCMA region was evaluated as a function of current density and anisotropy. As shown in Fig. \ref{fig:anisotropy_analysis}(a), skyrmions are unable to pass through regions of differing anisotropy unless sufficient current is provided. When the VCMA region anisotropy is higher than the remainder of the track, skyrmions are unable to overcome the anisotropy barrier to enter the VCMA region and are pinned outside; when the VCMA region anisotropy is lower than the remainder of the track, skyrmions enter the VCMA region but are unable to traverse the barrier to leave the VCMA region and are pinned inside.

\begin{figure}[!t]
\vspace{-1em}
\includegraphics[trim = 0 0 0 0, clip, width=0.75\columnwidth]{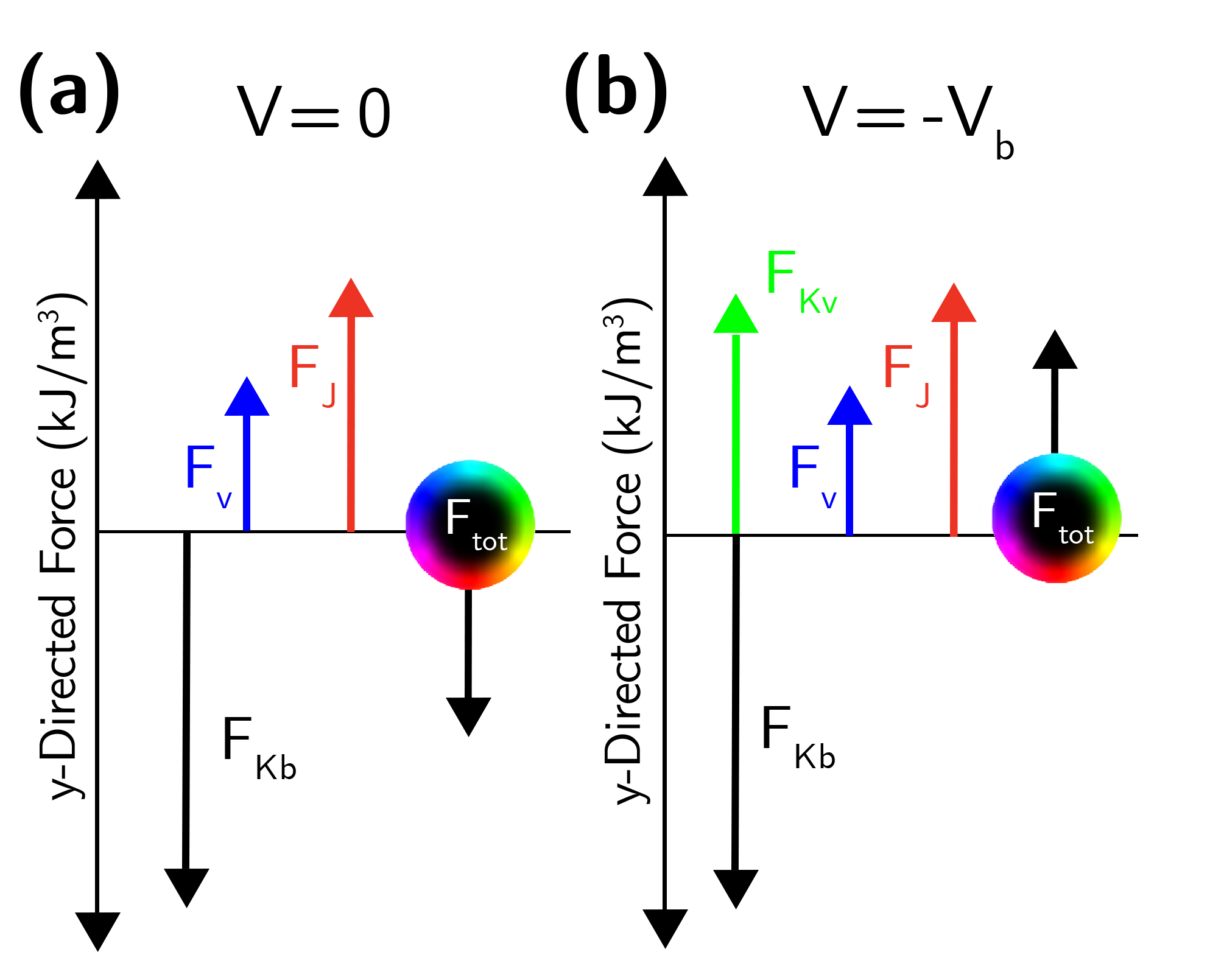}
\vspace{-1em}
\caption{Depiction of forces applied to skyrmion at VCMA region, with forces in the $+y$ direction assisting the skyrmion to traverse the barrier. The force resulting from the current density $(F_J)$ and the effective velocity force $(F_v)$ are in the $+y$ direction, while the anisotropy barrier results in a $-y$-directed force $F_{Kb}$. (a) For an applied voltage of $V=0$, the skyrmion has negative net force ($F_{tot} < 0$) and will be pinned by the anisotropy barrier. (b) When a voltage of $V=-V_b$ is applied, the anisotropy barrier is lowered by a $+y$-directed force $F_{Kv}$, leading to a positive net force ($F_{tot} > 0$) that enables the skyrmion to traverse the potential barrier.}
\label{fig:forces}
\end{figure}

It is noteworthy that Fig. \ref{fig:anisotropy_analysis}(a) is asymmetric about $K_{ub}=K_{u}$, as skyrmions with sufficient current are able to overcome larger energy barriers when leaving the VCMA region than when entering it. This is a result of the acceleration of the skyrmions as they propagate through the VCMA region, resulting in skyrmions attempting to leave the VCMA region with significantly greater velocity than when they entered the VCMA region.

This impact of velocity can be more directly observed when comparing the ability of stationary and moving skyrmions to traverse the VCMA region: all of the simulations in Fig. \ref{fig:anisotropy_analysis}(a) were performed for both stationary skyrmions and skyrmions moving at their terminal velocity, and in one case the moving skyrmion is able to traverse the VCMA region whereas the initially stationary skyrmion is not.

\begin{figure*}[!t]
\centering
\includegraphics[width=1.5\columnwidth]{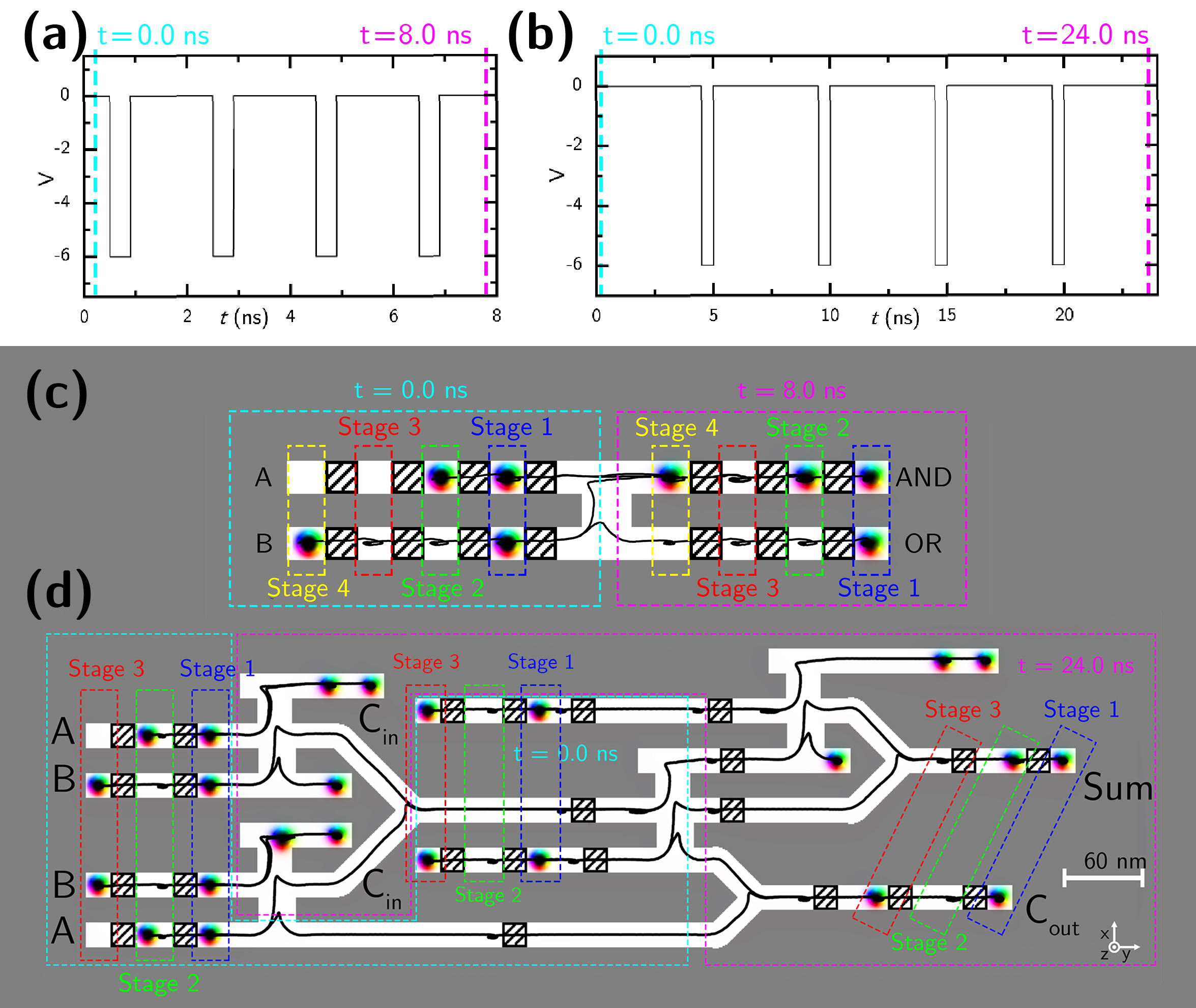}
\caption{VCMA synchronization of pipelined skyrmion logic circuits, with a constant current ($5\times 10^{10} {J}/{m^2}$) injected in the $+y$ direction. Clock waveforms are applied to the VCMA synchronizers in the (a) pipelined AND/OR gate and (b) pipelined one-bit full adder. Micromagnetic simulation results indicate the initial and final skyrmion states for the (c) pipelined AND/OR gate and (d) pipelined one-bit full adder, demonstrating proper logical operation.\vspace{-1em}}
\label{fig:verification_sims}
\end{figure*}

\subsection{Skyrmion Dynamics}

To further analyze the impact of skyrmion velocity, skyrmions were accelerated along a racetrack to determine the minimum pinning magnetic anisotropy as a function of current density. As shown in Fig. \ref{fig:anisotropy_analysis}(b), increases in current density necessitate larger anisotropy barriers to pin the skyrmions. Furthermore, skyrmions that enter the VCMA region with zero velocity are clearly observed to overcome less of an energy barrier than skyrmions that enter the VCMA region with their terminal velocity.

The impact of skyrmion velocity on the ability of a skyrmion to traverse an anisotropy barrier is further explored in Fig. \ref{fig:anisotropy_analysis}(c). This difference in pinning anisotropy is quadratic with respect to velocity, and it represents the contribution of the skyrmion velocity to the minimum pinning anisotropy. The cause of this increase can better understood by analyzing the Thiele equation (\ref{eq:2}) and its solution (\ref{eq:3}):\cite{Diaz2019}

\begin{equation}
    \bm{G}\times \bm{v}-\alpha \bm{\hat{D}v}+\bm{F}=0
    \label{eq:2}
\end{equation}
\begin{equation}
    \begin{pmatrix} 
    v_x \\
    v_y \\
    \end{pmatrix}
    = \frac{1}{\alpha^2 D^2 + G^2}
    \begin{pmatrix} 
    \alpha D F_x + F_y G \\
    \alpha D F_y - F_x G
    \end{pmatrix}
    \label{eq:3}
\end{equation}
\\
Comparing the steady state solution ($V_x = 0$) to the dynamic solution ($F_x=0$), there is a ratio between steady state velocity ($v_{ss}$) and dynamic velocity ($v_d$) equal to $1 + \frac{G^2}{\alpha^2 + D^2}$. These differences in velocity can be accounted for with an additional fictitious force in the $+y$ direction equal to $\frac{G^2}{\alpha^2 + D^2}V_d$, referred to as effective velocity force ($P$). Both the current density driving force ($J$) and the effective velocity force ($P$) contribute to the total effective driving force of the skyrmion ($F_{tot}$), and can be quantized in terms of anisotropy increase from Fig. \ref{fig:anisotropy_analysis}(c).

Figure \ref{fig:forces} depicts the impact of the VCMA clock pulse on the forces involved in the pinning interaction. In the absence of a clock pulse, the total force is negative, implying a transverse force in opposition to traversing the barrier; when a clock pulse is applied, the total force is positive, implying a transverse force that enables the skyrmion to traverse the barrier. To minimize the power dissipation and error rate of a skyrmion logic system, the VCMA parameters should be optimized to provide maximum reliability with minimal VCMA voltage modulation.

\section{Skyrmion Logic Circuits via VCMA Synchronization}

The VCMA clocking can be used to synchronize the motion of all the skyrmions within a system, thereby enabling the proper functionality of large-scale cascaded circuits. To prove the ability of the proposed VCMA synchronizers to enable cascaded logic gates, their functionality has been demonstrated in pipelined AND/OR and full adder circuits via micromagnetic simulation.

\subsection{Pipelined AND/OR Gate}

The effectiveness of the VCMA synchronizers can be demonstrated by adding pipelining to the simple AND/OR logic gate of Supplemental Material Fig. S1. As shown in Fig. \ref{fig:verification_sims}(a), \ref{fig:verification_sims}(c), and Video. 2, the skyrmions can be made to advance past one VCMA region during each 2 ns clock cycle, thereby providing the synchronization necessary for proper repulsion within the AND/OR gate junction. The logic gate has a throughput of one computation per clock cycle of both the AND and OR functions, and thus correctly performs computations for all four AND/OR conditions during this 8 ns simulation.\vspace{-1em}

\subsection{Pipelined One-Bit Full Adder}

To ensure the proper functionality of cascaded skyrmion logic gates, it is critical that the skyrmions entering each logic gate junction are synchronized. As shown in Fig. \ref{fig:verification_sims}(b), \ref{fig:verification_sims}(d), and Video. 3, the careful addition of VCMA synchronizers to a one-bit full adder circuit ensures proper skyrmion-skyrmion interactions despite the differing lengths of the skyrmion trajectories. In particular, the VCMA synchronizers are placed such that the skyrmions arrive contemporaneously at the logic gate junctions. Like the pipelined AND/OR gate, this full adder circuit therefore produces a throughput of one full addition computation (Sum and C$_{out}$) per clock cycle (5 ns). As there are two VCMA synchronizers within the full adder computation path, the latency of this full adder is two clock cycles (10 ns).

\section{Conclusion}

Reversible skyrmion logic provides a potential route to compute with less energy dissipation than Landauer's limit to the minimum energy for computation. In light of the importance of skyrmion synchronization to ensure proper skyrmion-skyrmion interactions within the logic gate junctions, this paper proposes skyrmion synchronization using VCMA modulation. We demonstrate the functionality and analyze the behavior of these VCMA synchronizers via micromagnetic simulation, providing a deeper understanding of their behavior that enables large-scale cascaded skyrmion circuit design. The use of VCMA synchronizers provides a reduced error rate and requires less energy than can be achieved with notch synchronizers,\cite{Chauwin2019} thereby greatly advancing the prospects for reversible computing with magnetic skyrmions.

\begin{DA}
	The data that support the findings of this study are available from the corresponding author upon reasonable request.
\end{DA}

\begin{acknowledgments}
	The authors thank E. Laws, J. McConnell, N. Nazir, and L. Philoon for technical support. This research is sponsored in part by the Texas Analog Center of Excellence Internship Program.
\end{acknowledgments}

\providecommand{\noopsort}[1]{}\providecommand{\singleletter}[1]{#1}%
%


\end{document}